\documentstyle[amstex, preprint, eqsecnum, aps]{revtex}

\tighten

\begin{document}
\title{Exact solution of the Hu-Paz-Zhang master equation}
\author{G. W. Ford\cite{bill} and R. F. O'Connell\cite{bob}}
\address{School of Theoretical Physics\\
Dublin Institute for Advanced Studies\\
10 Burlington Road\\
Dublin 4, Ireland}
\date{\today }
\maketitle

\begin{abstract}
The Hu-Paz-Zhang equation is a master equation for an oscillator coupled to
a linear passive bath. It is exact within the assumption that the oscillator
and bath are initially uncoupled \ Here an exact general solution is
obtained in the form of an expression for the Wigner function at time $t$ in
terms of the initial Wigner function. The result is applied to the motion of
a Gaussian wave packet and to that of a pair of such wave packets. A serious
divergence arising from the assumption of an initially uncoupled state is
found to be due to the zero-point oscillations of the bath and not removed
in a cutoff model. As a consequence, worthwhile results for the equation can
only be obtained in the high temperature limit, where zero-point
oscillations are neglected. In that limit closed form expressions for wave
packet spreading and attenuation of coherence are obtained. These results
agree within a numerical factor with those appearing in the literature,
which apply for the case of a particle at zero temperature that is suddenly
coupled to a bath at high temperature. On the other hand very different
results are obtained for the physically consistent case in which the initial
particle temperature is arranged to coincide with that of the bath.
\end{abstract}


\narrowtext

\section{Introduction}

The Hu-Paz-Zhang equation is a master equation with time-dependent
coefficients for a harmonic oscillator interacting with a linear passive
heat bath of oscillators. The equation is exact and general within the
assumption that in the initial state the bath is in equilibrium and not
coupled to the oscillator. It was first derived in generality using path
integral methods by Hu, Paz and Zhang \cite{hpz}, although equivalent
equations had been obtained earlier for the case of an Ohmic bath \cite
{leggett,unruh89}. See also \cite{karrlein}. Later a derivation in the form
of an equation for the Wigner function was given by Halliwell and Yu \cite
{halliwell}, who corrected a misprint in the earlier publication. Using the
notation of these last authors, the equation has the form 
\begin{eqnarray}
\frac{\partial W}{\partial t} &=&-\frac{1}{m}p\frac{\partial W}{\partial q}%
+m\Omega ^{2}(t)q\frac{\partial W}{\partial p}  \nonumber \\
&&+2\Gamma (t)\frac{\partial pW}{\partial p}+\hbar m\Gamma (t)h(t)\frac{%
\partial ^{2}W}{\partial p^{2}}+\hbar \Gamma (t)f(t)\frac{\partial ^{2}W}{%
\partial q\partial p},  \label{1.1}
\end{eqnarray}
where $\Omega ^{2}(t)$, $2\Gamma (t)$, $h(t)$, and $f(t)$ are time-dependent
parameters for which one has explicit expressions (see equations (\ref{2.19}%
) and (\ref{3.8}) below).

The integration of this equation, with its time-dependent coefficients,
appears to be a formidable problem. Indeed, earlier discussions have
generally used numerical methods. Our purpose here is, first of all, to
present an exact general solution of this equation. This solution can be
evaluated in explicit, closed form for many problems of interest. In
particular, we exhibit the solution for two such problems: an initial state
corresponding to a Gaussian minimum uncertainty wave packet and an initial
state corresponding to a widely separated pair of such wave packets. We use
these results to accomplish our second purpose, which is to critically
examine the assumption of an uncoupled initial state. We find that there is
a serious difficulty arising from this assumption: the zero-point
oscillations of the bath give rise to a divergence that leads to an
instantaneous spread of a wave packet to infinite width. In effect, the
state instantaneously disappears! The result is that meaningful results can
be obtained only in the high temperature limit, where one conventionally
neglects the zero-point oscillations. Even in this limit, we find
significant difficulties arising from the fact that translational invariance
is broken. Nevertheless, we find for short times expressions for wave packet
spreading and attenuation of coherence that are consistent with those found
by earlier authors. On the other hand, by adjusting the initial temperature
of the particle to be the same as that of the bath, we find in the high
temperature Ohmic limit results consistent with exact calculations which
take into account entanglement at all times \cite{ford01a,ford01b}.

The plan of this paper is as follows. The basis for our discussion is the
quantum Langevin equation, so we begin in Section II with a description of
that equation and its solution, first for the stationary case, then for the
initial value case, and finally for the form local in time. \ We next in a
short Section III we give a simple derivation of the exact master equation (%
\ref{1.1}). Then in Section IV we derive our general solution. The key
result, given in (\ref{4.15}), is an explicit expression for the Wigner
function at time $t$ in terms of the initial Wigner function. A particularly
useful result, given in (\ref{4.24}), is an expression for the probability
distribution at time $t$. In Section V we first evaluate this expression to
find the probability distribution corresponding to an initial Gaussian wave
packet. There we find the divergence mentioned above, which in Appendix A is
shown to be present even in a model with a high frequency cutoff. Also in
section V we consider the motion of a pair of Gaussian wave packets (Schr%
\"{o}dinger ``cat'' state) and obtain an explicit expression for the
attenuation of coherence. For the case of a particle at temperature zero
suddenly coupled to a bath at high temperature $T$, this leads to an
expression for the decoherence time equivalent with that appearing in many
places in the literature. But for a consistent initial state in which the
temperature of the particle is adjusted to coincide with that of the bath, a
very different expression for the decoherence time is obtained, an
expression corresponding to decoherence without dissipation \cite{ford01b}.
Finally, in Section VI we summarize our results and make some concluding
remarks.

\section{The Langevin equation}

\subsection{Stationary process}

The Langevin equation is a Heisenberg equation of motion for $x(t)$, the
dynamical variable corresponding to the coordinate of a Brownian particle in
equilibrium with a linear passive heat bath. For the case of a particle in
an external oscillator potential, this equation for the stationary process
has the well known form \cite{flo1988}, 
\begin{equation}
m\ddot{x}+\int_{-\infty }^{t}dt^{\prime }\mu (t-t^{\prime })\dot{x}%
(t^{\prime })+Kx=F(t),  \label{2.1}
\end{equation}
where $\mu (t)$ is the memory function, $K$ is the oscillator force constant
and $F(t)$ is a fluctuating operator force with mean $\left\langle
F(t)\right\rangle =0$, and whose correlation and commutator are given by 
\begin{eqnarray}
\frac{1}{2}\left\langle F(t^{\prime })F(t)+F(t)F(t^{\prime })\right\rangle
&=&\frac{1}{\pi }\int_{0}^{\infty }d\omega {\rm Re}\{\tilde{\mu}(\omega
+i0^{+})\}\hbar \omega \coth \frac{\hbar \omega }{2kT}\cos \omega
(t-t^{\prime }),  \nonumber \\
\lbrack F(t),F(t^{\prime })] &=&\frac{2\hbar }{i\pi }\int_{0}^{\infty
}d\omega {\rm Re}\{\tilde{\mu}(\omega +i0^{+})\}\omega \sin \omega
(t-t^{\prime }).  \label{2.2}
\end{eqnarray}
Here $\tilde{\mu}(z)$ is the Fourier transform of the memory function, 
\begin{equation}
\tilde{\mu}(z)=\int_{0}^{\infty }dt\mu (t)e^{izt}.  \label{2.3}
\end{equation}
As a consequence of the second law of thermodynamics, $\tilde{\mu}(z)$ must
be what is called a positive real function: analytic and with real part
positive in the upper half plane. In particular, $\tilde{\mu}(z)$ can be
represented in terms of the real part of its boundary value on the real axis
through the Stieltjes inversion theorem 
\begin{equation}
\tilde{\mu}(z)=-icz+{\frac{2iz}{\pi }}\int_{0}^{\infty }d\omega {\frac{{\rm %
Re}\{\tilde{\mu}(\omega +i0^{+})\}}{z^{2}-\omega ^{2}},}  \label{2.4}
\end{equation}
where $c$ is a positive constant.

The solution of the Langevin equation (\ref{2.1}) can be written 
\begin{equation}
x_{{\rm s}}(t)=\int_{-\infty }^{t}dt^{\prime }G(t-t^{\prime })F(t^{\prime }),
\label{2.5}
\end{equation}
where $G(t)$, the Green function, is given by 
\begin{equation}
G(t)=\frac{1}{2\pi }\int_{-\infty }^{\infty }d\omega \alpha (\omega
+i0^{+})e^{-i\omega t},  \label{2.6}
\end{equation}
with $\alpha (z)$ the familiar response function 
\begin{equation}
\alpha (z)=\frac{1}{-mz^{2}-iz\tilde{\mu}(z)+K}.  \label{2.7}
\end{equation}
Here we have introduced a subscript s to emphasize that $x_{{\rm s}}(t)$ is
a stationary operator-process, in the sense that correlations, probability
distributions, etc. for this dynamical variable are invariant under
time-translation ($t\rightarrow t+t_{0}$). In particular, the correlation, 
\begin{equation}
\frac{1}{2}\left\langle x_{{\rm s}}(t)x_{{\rm s}}(t^{\prime })+x_{{\rm s}%
}(t^{\prime })x_{{\rm s}}(t)\right\rangle =\frac{\hbar }{\pi }%
\int_{0}^{\infty }d\omega {\rm Im}\{\alpha (\omega +i0^{+})\}\coth \frac{%
\hbar \omega }{2kT}\cos \omega (t-t^{\prime }),  \label{2.8}
\end{equation}
is a function only of the time-difference $t-t^{\prime }$. In addition, for
the free particle, where $K=0$, the process is invariant under space
translation ($x\rightarrow x+a$).

\subsection{Langevin equation for the initial value problem}

The description of the system given by the Langevin equation can be realized
by a bath of harmonic oscillators. Perhaps the simplest such system, and the
one we use as the basis of our discussion of the Hu-Paz-Zhang equation and
its solution, is the independent oscillator model, for which the Hamiltonian
is \cite{zwanzig,flo1988} 
\begin{equation}
H=\frac{p^{2}}{2m}+\frac{1}{2}Kx^{2}+\sum_{j}\{\frac{p_{j}^{2}}{2m_{j}}+%
\frac{1}{2}m_{j}\omega _{j}^{2}(q_{j}-x)^{2}\}.  \label{2.9}
\end{equation}
Writing the equations of motion and then eliminating the bath variables in
terms of their initial values, one obtains the Langevin equation for the
oscillator with given initial values \cite{kac}, 
\begin{equation}
m\ddot{x}+\int_{0}^{t}dt^{\prime }\mu (t-t^{\prime })\dot{x}(t^{\prime
})+Kx=-\mu (t)x(0)+F(t),  \label{2.10}
\end{equation}
where the memory function is given by 
\begin{equation}
\mu (t)=\sum_{j}m_{j}\omega _{j}^{2}\cos \omega _{j}t\theta (t),
\label{2.11}
\end{equation}
while the random force is given in terms of the initial bath variables by 
\begin{equation}
F(t)=\sum_{j}\{q_{j}(0)m_{j}\omega _{j}^{2}\cos \omega _{j}t+p_{j}(0)\omega
_{j}\sin \omega _{j}t\}.  \label{2.12}
\end{equation}

To express the solution of this equation, we first note that the Green
function (\ref{2.6}) vanishes for negative times and for positive times is
the solution of the homogeneous equation, 
\begin{equation}
m\ddot{G}+\int_{0}^{t}dt^{\prime }\mu (t-t^{\prime })\dot{G}(t^{\prime
})+KG=0,  \label{2.13}
\end{equation}
with the initial conditions 
\begin{equation}
G(0)=0,\qquad \dot{G}(0)=\frac{1}{m}.  \label{2.14}
\end{equation}
With this, we can show that the general solution of the initial value
Langevin equation (\ref{2.10}) is given by 
\begin{eqnarray}
x(t) &=&m\dot{G}(t)x(0)+mG(t)\dot{x}(0)+X(t)  \nonumber \\
\dot{x}(t) &=&m\ddot{G}(t)x(0)+m\dot{G}(t)\dot{x}(0)+\dot{X}(t),
\label{2.15}
\end{eqnarray}
where we have introduced the fluctuating position operator, 
\begin{equation}
X(t)=\int_{0}^{t}dt^{\prime }G(t-t^{\prime })F(t^{\prime }).  \label{2.16}
\end{equation}

In our subsequent discussion we assume that at $t=0$ the system is in a
state in which the oscillator is not coupled to the bath and that the bath
is in equilibrium at temperature $T$. In particular this means that the
initial coordinates of the oscillator are not correlated with those of the
bath, i.e., $\left\langle x(0)F(t)\right\rangle =\left\langle \dot{x}%
(0)F(t)\right\rangle =0$. On the other hand, with regard to the bath, the
equilibrium is with respect to the bath Hamiltonian, $H_{{\rm bath}}$,
obtained by setting the oscillator variables $x$ and $p$ equal to zero in (%
\ref{2.9}), 
\begin{equation}
H_{{\rm bath}}=\sum_{j}(\frac{p_{j}^{2}}{2m_{j}}+\frac{1}{2}m_{j}\omega
_{j}^{2}q_{j}^{2}).  \label{2.17}
\end{equation}
With this we find $\left\langle F(t)\right\rangle =0$, and the correlation
and commutator are the same as those for the stationary equation, given in (%
\ref{2.2}).

Typically, the memory function $\mu (t)$ falls to zero in a very short time $%
\tau $, called the relaxation time of the bath. For times long compared with
this bath relaxation time, the extra term on the right hand side of (\ref
{2.10}) vanishes, but only for much longer times, times long compared with
the oscillator decay time, will this equation become the stationary
equation, with the lower limit on the integration taken to be $-\infty $. To
be more specific, we note from the general expression (\ref{2.10}) for the
Green function that, so long as the oscillator force constant $K$ is not
zero, the Green function will vanish exponentially for long times. This
follows from the Tauberian theorem: the asymptotic behavior of a function
depends upon the low frequency behavior of its Fourier transform. It follows
that, for long times, the dependence upon the initial coordinates in (\ref
{2.15}) disappears and, from a comparison of (\ref{2.16}) with the
expression (\ref{2.5}) for $x_{{\rm s}}(t)$, that $X(t)$ becomes the
solution of the stationary Langevin equation (\ref{2.1}).

\subsection{Form local in time}

We want now to write the Langevin equation (\ref{2.10}) in the form of an
equation that is local in time with time-dependent coefficients. To get this
form, we first invert the equations (\ref{2.15}) to express the initial
variables in terms of those at time $t$. We next form the time derivative of
the second of the equations (\ref{2.15}) and then insert these expressions
for the initial variables in the right hand side. We can write the result in
the form 
\begin{equation}
\ddot{x}+2\Gamma (t)\dot{x}+\Omega ^{2}(t)x=\frac{1}{m}F(t),  \label{2.18}
\end{equation}
where we have introduced the quantities 
\begin{eqnarray}
2\Gamma (t) &=&\frac{G(t)\dddot{G}(t)-\dot{G}(t)\ddot{G}(t)}{\dot{G}%
^{2}(t)-G(t)\ddot{G}(t)}=-\frac{d\log (\dot{G}^{2}-G\ddot{G})}{dt}, 
\nonumber \\
\Omega ^{2}(t) &=&\frac{\ddot{G}^{2}(t)-\dot{G}(t)\dddot{G}(t)}{\dot{G}%
^{2}(t)-G(t)\ddot{G}(t)}.  \label{2.19}
\end{eqnarray}
This equation is the local form we seek. In obtaining this form we have used
the fact that $X(t)$ is the solution of the inhomogeneous equation (\ref
{2.10}) with the initial conditions $X(0)=0$ and $\dot{X}(0)=0$. It follows
that $X(t)$ is also the solution of the local equation (\ref{2.18}) with the
same initial conditions. The fact that the symbols used for the quantities (%
\ref{2.19}) also appear in the exact master equation (\ref{1.1}) is not
accidental, as we shall see in the next section, where we give a derivation
of that equation.

\section{The time-dependent master equation}

The strategy for deriving the exact master equation is to obtain expressions
for the first and second moments, first from the equation (\ref{1.1}) and
then from the local Langevin equation (\ref{2.18}). From a comparison, we
obtain explicit expressions for the time-dependent parameters in the
Hu-Paz-Zhang equation (\ref{1.1}).

In forming the moments of the equation (\ref{1.1}), we take note of the
relations \cite{hillery} 
\begin{eqnarray}
x\rho &\leftrightarrow &(q+\frac{i\hbar }{2}\frac{\partial }{\partial p})W,
\qquad \rho x\leftrightarrow (q-\frac{i\hbar }{2}\frac{\partial }{\partial p}%
)W,  \nonumber \\
p\rho &\leftrightarrow &(p-\frac{i\hbar }{2}\frac{\partial }{\partial q})W,
\qquad \rho p\leftrightarrow (p+\frac{i\hbar }{2}\frac{\partial }{\partial q}%
)W.  \label{3.1}
\end{eqnarray}
Here on the left $x$ and $p$ are the position and momentum operators for the
oscillator, while $\rho $ is the density matrix operator. On the right, $q$
and $p$ are the c-number variables of the Wigner function $W(q, p;t\dot{)}$.
Thus, for example, 
\begin{equation}
\left\langle x\right\rangle \equiv {\rm Tr}\{x\rho \}=\int_{-\infty
}^{\infty }dq\int_{-\infty }^{\infty }dp(q+\frac{i\hbar }{2}\frac{\partial }{%
\partial p})W.  \label{3.2}
\end{equation}
In this way, forming the first moments of equation (\ref{1.1}), we find 
\begin{equation}
\left\langle \dot{x}\right\rangle =\frac{1}{m}\left\langle p\right\rangle ,
\qquad \left\langle \dot{p}\right\rangle =-2\Gamma (t)\left\langle
p\right\rangle -m\Omega ^{2}(t)\left\langle x\right\rangle .  \label{3.3}
\end{equation}
Eliminating $\left\langle p\right\rangle $, we find 
\begin{equation}
\left\langle \ddot{x}\right\rangle +2\Gamma (t)\left\langle \dot{x}%
\right\rangle +\Omega ^{2}(t)\left\langle x\right\rangle =0.  \label{3.4}
\end{equation}
If now, we form the mean of the local Langevin equation (\ref{2.18}), using
the fact that $\left\langle X(t)\right\rangle =0$, we get exactly the same
equation, but with now the quantities $2\Gamma (t)$ and $\Omega ^{2}(t)$
given by the expressions (\ref{2.19}).

Next, forming the second moments of the equation (\ref{1.1}), we find 
\begin{eqnarray}
\frac{d\left\langle x^{2}\right\rangle }{dt} &=&\frac{1}{m}\left\langle
xp+px\right\rangle ,  \nonumber \\
\frac{d\left\langle xp+px\right\rangle }{dt} &=&\frac{2}{m}\left\langle
p^{2}\right\rangle -2m\Omega ^{2}\left\langle x^{2}\right\rangle -2\Gamma
\left\langle xp+px\right\rangle +2\hbar \Gamma f,  \nonumber \\
\frac{d\left\langle p^{2}\right\rangle }{dt} &=&-m\Omega ^{2}\left\langle
xp+px\right\rangle -4\Gamma \left\langle p^{2}\right\rangle +2\hbar m\Gamma
h.  \label{3.5}
\end{eqnarray}
On the other hand, using the local Langevin equation (\ref{2.18}), we find 
\begin{eqnarray}
\frac{d\left\langle x^{2}\right\rangle }{dt} &=&\left\langle x\dot{x}+\dot{x}%
x\right\rangle ,  \nonumber \\
\frac{d\left\langle x\dot{x}+\dot{x}x\right\rangle }{dt} &=&2\left\langle 
\dot{x}^{2}\right\rangle +\left\langle x\ddot{x}+\ddot{x}x\right\rangle 
\nonumber \\
&=&2\left\langle \dot{x}^{2}\right\rangle -2\Omega ^{2}(t)\left\langle
x^{2}\right\rangle -2\Gamma (t)\left\langle x\dot{x}+\dot{x}x\right\rangle 
\nonumber \\
&&+\frac{1}{m}\left\langle x(t)F(t)+F(t)x(t)\right\rangle ,  \nonumber \\
\frac{d\left\langle \dot{x}^{2}\right\rangle }{dt} &=&\left\langle \dot{x}%
\ddot{x}+\ddot{x}\dot{x}\right\rangle  \nonumber \\
&=&-\Omega ^{2}(t)\left\langle x\dot{x}+\dot{x}x\right\rangle -4\Gamma
(t)\left\langle \dot{x}^{2}\right\rangle  \nonumber \\
&&+\frac{1}{m}\left\langle \dot{x}(t)F(t)+F(t)\dot{x}(t)\right\rangle .
\label{3.6}
\end{eqnarray}
Now, in the right hand side of the second of these equations we use the fact 
$x(t)-X(t)$ is not correlated with $F(t)$ \ This should be clear since, as
we see from (\ref{2.15}), this combination depends only upon the initial
coordinates of the oscillator. Therefore we can replace $x(t)$ with $X(t)$
in the correlations of $x$ with $F$. Using the same argument in the last of
these equations, we see that after a little rearrangement we can write 
\begin{eqnarray}
\frac{d\left\langle x^{2}\right\rangle }{dt} &=&\left\langle x\dot{x}+\dot{x}%
x\right\rangle ,  \nonumber \\
\frac{d\left\langle x\dot{x}+\dot{x}x\right\rangle }{dt} &=&2\left\langle 
\dot{x}^{2}\right\rangle -2\Omega ^{2}(t)\left\langle x^{2}\right\rangle
-2\Gamma (t)\left\langle x\dot{x}+\dot{x}x\right\rangle  \nonumber \\
&&+\frac{1}{m}\left\langle X(t)F(t)+F(t)X(t)\right\rangle ,  \nonumber \\
\frac{d\left\langle \dot{x}^{2}\right\rangle }{dt} &=&-\Omega
^{2}(t)\left\langle x\dot{x}+\dot{x}x\right\rangle -4\Gamma (t)\left\langle 
\dot{x}^{2}\right\rangle  \nonumber \\
&&+\frac{1}{m}\left\langle \dot{X}(t)F(t)+F(t)\dot{X}(t)\right\rangle .
\label{3.7}
\end{eqnarray}

We now compare these equations with the equations (\ref{3.5}) obtained from
the equation (\ref{1.1}). In doing so we must interpret $p=m\dot{x}$. We see
then that we can identify 
\begin{eqnarray}
2\hbar \Gamma (t)f(t) &=&\left\langle X(t)F(t)+F(t)X(t)\right\rangle , 
\nonumber \\
2\hbar \Gamma (t)h(t) &=&\left\langle \dot{X}(t)F(t)+F(t)\dot{X}%
(t)\right\rangle .  \label{3.8}
\end{eqnarray}
This completes the derivation of the exact master equation (\ref{1.1}), with
explicit expressions for the time-dependent coefficients.

\section{General solution of the time-dependent master equation}

The task of solving the equation (\ref{1.1}), with its time-dependent
coefficients given by the complicated expressions (\ref{2.19}) and (\ref{3.8}%
), appears formidable. Indeed, if one were presented with the equation with
no idea of the origin of the coefficients its solution would be very
difficult. But we have in (\ref{2.15}) an explicit solution of Langevin
equation describing the underlying motion. This will allow us to construct
the general solution of the equation.

To begin, we remind ourselves (\ref{1.1}) is an equation for the reduced
density matrix, given by the partial trace over the bath coordinates. That
is, 
\begin{equation}
W(q,p;t)=\int d{\bf q}\int d{\bf p}W_{{\rm system}}(q,p;{\bf q},{\bf p};t).
\label{4.1}
\end{equation}
Here $W_{{\rm system}}$ is the Wigner function for the system of oscillator
and bath, with ${\bf q}=(q_{1},q_{2},\cdots )$ and ${\bf p}%
=(p_{1},p_{2},\cdots )$ the bath coordinates and momenta. Now, the system is
one of coupled oscillators and for such a system the solution of the
equation of motion is formally identical to that for the corresponding
classical system. That is, the Wigner function for the system at time $t$ is
related to that at time $t=0$ through the relation 
\begin{equation}
W_{{\rm system}}(q,p;{\bf q},{\bf p};t)=W_{{\rm system}}(q(0),p(0);{\bf q}%
(0),{\bf p}(0);0),  \label{4.2}
\end{equation}
where $q(0),p(0);{\bf q}(0),{\bf p}(0)$ are the initial values for which the
solution of the equations of motion is such that $q(t)=q$, $p(t)=p$, ${\bf q}%
(t)={\bf q}$, ${\bf p}(t)={\bf p}$. Finally, we remind ourselves that for
the Hu-Paz-Zhang equation the initial state is a product state corresponding
to a Wigner function of the form \cite{hillery} 
\begin{equation}
W_{{\rm system}}(q,p;{\bf q},{\bf p};0)=W(q,p;0)\prod_{j}w_{j}(q_{j},p_{j}).
\label{4.3}
\end{equation}
Here, on the right $W(q,p;0)$ is the initial Wigner function for the
oscillator and the product is the Wigner function for the bath, in which $%
w_{j}(q_{j},p_{j})$ is the Wigner function for a single oscillator of mass $%
m_{j}$ and frequency $\omega _{j}$, 
\begin{equation}
w_{j}(q_{j},p_{j})=\frac{1}{\pi \hbar \coth (\hbar \omega _{j}/2kT)}\exp
\left\{ -\frac{p_{j}^{2}+m_{j}^{2}\omega _{j}^{2}q_{j}^{2}}{m_{j}\hbar
\omega _{j}\coth (\hbar \omega _{j}/2kT)}\right\} .  \label{4.4}
\end{equation}
Combining these results, we see that the reduced density matrix at time $t$
is given by 
\begin{equation}
W(q,p;t)=\int d{\bf q}\int d{\bf p}W(q(0),p(0);0)%
\prod_{j}w_{j}(q_{j}(0),p_{j}(0)).  \label{4.5}
\end{equation}
This reduced density matrix is the solution that we seek. So far, however,
all we have done is to carefully indicate the definition of this quantity,
we must now carry out the indicated operations to obtain an explicit
expression.

As a first step we transform the integration to the initial bath
coordinates, holding $q$ and $p$ fixed. Under this transformation, 
\begin{equation}
d{\bf q}d{\bf p}=\frac{\partial (q,p;{\bf q},{\bf p})}{\partial (q,p;{\bf q}%
(0),{\bf p}(0))}d{\bf q}(0)d{\bf p}(0),  \label{4.6}
\end{equation}
where the factor is the Jacobean of the transformation, for which we have
used the notation of Landau and Lifshitz \cite{landau58}. But, 
\begin{eqnarray}
\frac{\partial (q,p;{\bf q},{\bf p})}{\partial (q,p;{\bf q}(0),{\bf p}(0))}
&=&\frac{\partial (q,p;{\bf q},{\bf p})}{\partial (q(0),p(0);{\bf q}(0),{\bf %
p}(0))}\frac{\partial (q(0),p(0);{\bf q}(0),{\bf p}(0))}{\partial (q,p;{\bf q%
}(0),{\bf p}(0))}  \nonumber \\
&=&\frac{\partial (q(0),p(0);{\bf q}(0),{\bf p}(0))}{\partial (q,p;{\bf q}%
(0),{\bf p}(0))}  \nonumber \\
&=&\left( \frac{\partial (q,p;{\bf q}(0),{\bf p}(0))}{\partial (q(0),p(0);%
{\bf q}(0),{\bf p}(0))}\right) ^{-1}  \nonumber \\
&=&\left( \frac{\partial q}{\partial q(0)}\frac{\partial p}{\partial p(0)}-%
\frac{\partial q}{\partial p(0)}\frac{\partial p}{\partial q(0)}\right)
^{-1}.  \label{4.7}
\end{eqnarray}
Here, we use in the first line the fact that the Jacobean of two successive
transformation is the product of the Jacobians, in the second line the fact
that the motion of the system corresponds to a canonical transformation for
which the Jacobean is unity, in the third line the fact that the Jacobean of
the inverse transformation is the reciprocal of that of the direct
transformation, and finally in the last line the definition of the Jacobean
as the determinant of the matrix of partial derivatives. Now, to evaluate
this Jacobean we use the solution (\ref{2.15}), which we write in the form 
\begin{eqnarray}
q &\equiv &q(t)=m\dot{G}(t)q(0)+G(t)p(0)+X(t),  \nonumber \\
p &\equiv &p(t)=m^{2}\ddot{G}(t)q(0)+m\dot{G}(t)p(0)+m\dot{X}(t).
\label{4.8}
\end{eqnarray}
Here we recall, from the definitions (\ref{2.16}) of $X(t)$ and the
expression (\ref{2.12}) for $F(t)$ that $X(t)$ depends only on the initial
coordinates of the bath, which are held fixed in forming the partial
derivatives in the last line of (\ref{4.7}). Therefore, we see that 
\begin{equation}
\frac{\partial (q,p;{\bf q},{\bf p})}{\partial (q,p;{\bf q}(0),{\bf p}(0))}=%
\frac{1}{m^{2}(\dot{G}^{2}-G\ddot{G})}  \label{4.9}
\end{equation}
and, using (\ref{4.6}) we can write (\ref{4.5}) in the form 
\begin{equation}
W(q,p;t)=\frac{\left\langle W(q(0),p(0);0)\right\rangle }{m^{2}(\dot{G}^{2}-G%
\ddot{G})},  \label{4.10}
\end{equation}
where the brackets represent the average over the initial equilibrium
distribution of the bath. Again, we remind ourselves that in the integrand $%
q(0)$ and $p(0)$ are obtained by inverting the equations (\ref{4.8}). That
is, 
\begin{eqnarray}
q(0) &=&\frac{m\dot{G}(q-X)-G(p-m\dot{X})}{m^{2}(\dot{G}^{2}-G\ddot{G})}, 
\nonumber \\
p(0) &=&\frac{-m^{2}\ddot{G}(q-X)+m\dot{G}(p-m\dot{X})}{m^{2}(\dot{G}^{2}-G%
\ddot{G})}.  \label{4.11}
\end{eqnarray}
Since $X$ is linear in the initial bath variables, its average has the
Gaussian property: averages of all moments can be expressed in terms of
those of the second moment.

We can carry out this average if we introduce the Fourier transform of the
initial Wigner function, writing 
\begin{equation}
W(q,p;0)=\frac{1}{(2\pi \hbar )^{2}}\int_{-\infty }^{\infty }dQ\int_{-\infty
}^{\infty }dP\tilde{W}(Q,P;0)e^{i(Pq+Qp)/\hbar }  \label{4.12}
\end{equation}
Inserting this in (\ref{4.10}), we can write 
\begin{equation}
W(q,p;t)=\frac{1}{(2\pi \hbar m)^{2}(\dot{G}^{2}-G\ddot{G})}\int_{-\infty
}^{\infty }dQ\int_{-\infty }^{\infty }dP\tilde{W}(Q,P;0)\left\langle \exp
\{i[Pq(0)+Qp(0)]/\hbar \}\right\rangle .  \label{4.13}
\end{equation}
The form of this result can be made a bit simpler if we introduce a
transformation to variables $r$ and $s$, defined by 
\begin{equation}
Q=m\dot{G}r+Gs,\qquad P=m^{2}\ddot{G}r+m\dot{G}s.  \label{4.14}
\end{equation}
We then find that $Pq(0)+Qp(0)=r(p-m\dot{X})+s(q-X)$ and $dQdP=m^{2}(\dot{G}%
^{2}-G\ddot{G})drds$, so that (\ref{4.13}) becomes 
\begin{eqnarray}
W(q,p;t) &=&\frac{1}{(2\pi \hbar )^{2}}\int_{-\infty }^{\infty
}dr\int_{-\infty }^{\infty }ds\tilde{W}(m\dot{G}r+Gs,m^{2}\ddot{G}r+m\dot{G}%
s;0)  \nonumber \\
&&\times e^{i(rp+sq)/\hbar }e^{-\frac{1}{2\hbar ^{2}}(m^{2}\left\langle \dot{%
X}^{2}\right\rangle r^{2}+m\left\langle X\dot{X}+\dot{X}X\right\rangle
rs+\left\langle X^{2}\right\rangle s^{2})},  \label{4.15}
\end{eqnarray}
where we have used the Gaussian property to write 
\begin{equation}
\left\langle e^{-i(m\dot{X}r+Xs)/\hbar }\right\rangle =e^{-\frac{1}{2\hbar
^{2}}(m^{2}\left\langle \dot{X}^{2}\right\rangle r^{2}+m\left\langle X\dot{X}%
+\dot{X}X\right\rangle rs+\left\langle X^{2}\right\rangle s^{2})}.
\label{4.16}
\end{equation}
This is the form of the solution that is perhaps most useful. In its
evaluation, the Green function $G(t)$ is given by (\ref{2.6}), while $X(t)$
is given by (\ref{2.16}) and its correlations are evaluated using (\ref{2.2}%
).

While we find (\ref{4.15}) to be the most useful form of the solution, one
can insert the inverse of the Fourier transform (\ref{4.12}) to express the
solution in the form of a transition operator acting on the initial Wigner
function, 
\begin{equation}
W(q,p;t)=\int_{-\infty }^{\infty }dq^{\prime }\int_{-\infty }^{\infty
}dp^{\prime }P(q,p;q^{\prime },p^{\prime };t)W(q^{\prime },p^{\prime };0).
\label{4.17}
\end{equation}
Here $P(q,p;q^{\prime },p^{\prime };t)$, called the transition probability,
can be written 
\begin{equation}
P(q,p;q^{\prime },p^{\prime };t)=\frac{1}{2\pi \sqrt{\det {\bf A}}}\exp \{-%
\frac{1}{2}{\bf R\cdot A}^{-1}{\bf \cdot R\},}  \label{4.18}
\end{equation}
where we have used a dyadic notation with 
\begin{eqnarray}
{\bf A}(t) &=&\left( 
\begin{array}{cc}
m^{2}\left\langle \dot{X}^{2}\right\rangle & \frac{m}{2}\left\langle X\dot{X}%
+\dot{X}X\right\rangle \\ 
\frac{m}{2}\left\langle X\dot{X}+\dot{X}X\right\rangle & \left\langle
X^{2}\right\rangle
\end{array}
\right) ,  \nonumber \\
{\bf R}(t) &=&\left( 
\begin{array}{c}
p-\left\langle p(t)\right\rangle \\ 
q-\left\langle q(t)\right\rangle
\end{array}
\right) .  \label{4.19}
\end{eqnarray}
Here, in ${\bf R}$, the quantities $\left\langle q(t)\right\rangle $ and $\
\left\langle p(t)\right\rangle $ correspond to the mean of the initial value
solution (\ref{2.15}) with initial values $q^{\prime }$ and $p^{\prime }$.
That is, 
\begin{eqnarray}
\left\langle q(t)\right\rangle &=&m\dot{G}(t)q^{\prime }+G(t)p^{\prime }, 
\nonumber \\
\left\langle p(t)\right\rangle &=&m^{2}\ddot{G}(t)q^{\prime }+m\dot{G}%
(t)p^{\prime }.  \label{4.20}
\end{eqnarray}

It is of interest that this expression (\ref{4.18}) for the transition
probability is formally the same as that for the classical Kramers equation 
\cite{risken}. The difference is that the Green function and the mean square
of the fluctuating position and velocity operators here are for a quantum
oscillator interacting with an arbitrary heat bath, while in the classical
solution of the Kramers equation they are for a classical oscillator
interacting with an Ohmic bath. Another significant difference is that the
solution (\ref{4.20}) is that of the mean of the initial value Langevin
equation (\ref{2.10}), with the term $-\mu (t)x(0)$ on the right hand side.
The classical solution of the Kramers equation corresponds to dropping this
term, since it is describing the classical stationary process.

As a first illustration of the utility of the form (\ref{4.15}), we show how
the equilibrium solution arises for long times. First we recall that, so
long as the oscillator force constant $K$ is not zero, the Green function
will vanish as $t\rightarrow \infty $. Next, we recall that, again as $%
t\rightarrow \infty $, $X(t)\rightarrow x_{{\rm s}}(t)$, the solution (\ref
{2.5}) of the stationary Langevin equation (\ref{2.1}). Thus, we see almost
by inspection that 
\begin{equation}
W(q,p;t)%
\mathrel{\mathop{\rightarrow }\limits_{t\rightarrow \infty }}%
\frac{1}{2\pi m\sqrt{\left\langle x_{{\rm s}}^{2}\right\rangle \left\langle 
\dot{x}_{{\rm s}}^{2}\right\rangle }}\exp \{-\frac{p^{2}}{2m^{2}\left\langle 
\dot{x}_{{\rm s}}^{2}\right\rangle }-\frac{q^{2}}{2\left\langle x_{{\rm s}%
}^{2}\right\rangle }\},  \label{4.21}
\end{equation}
where $\left\langle x_{{\rm s}}^{2}\right\rangle $ and $\left\langle \dot{x}%
_{{\rm s}}^{2}\right\rangle $ are the equilibrium values of the mean square
position and velocity, given by the fluctuation-dissipation theorem, 
\begin{eqnarray}
\left\langle x_{{\rm s}}^{2}\right\rangle &=&\frac{\hbar }{\pi }%
\int_{0}^{\infty }d\omega {\rm Im}\{\alpha (\omega +i0^{+})\}\coth \frac{%
\hbar \omega }{2kT},  \nonumber \\
\left\langle \dot{x}_{{\rm s}}^{2}\right\rangle &=&\frac{\hbar }{\pi }%
\int_{0}^{\infty }d\omega \omega ^{2}{\rm Im}\{\alpha (\omega
+i0^{+})\}\coth \frac{\hbar \omega }{2kT}.  \label{4.22}
\end{eqnarray}
This result is perhaps more familiar in the weak coupling limit, where ${\rm %
Im}\{\alpha (\omega +i0^{+})\}\rightarrow \pi \delta (\omega -\omega
_{0})/2m\omega _{0}$ with $\omega _{0}=\sqrt{K/m}$. Then $\left\langle \dot{x%
}_{{\rm s}}^{2}\right\rangle =\omega _{0}^{2}\left\langle x_{{\rm s}%
}^{2}\right\rangle =\frac{\hbar \omega _{0}}{2m}\coth \frac{\hbar \omega _{0}%
}{2kT}$ and (\ref{4.21}) becomes the familiar equilibrium form of the Wigner
function for the uncoupled oscillator\cite{hillery}.

Finally, we remark that the interest is most often in the probability
density at time $t$, given by 
\begin{equation}
P(x;t)=\int_{-\infty }^{\infty }dpW(x,p;t).  \label{4.23}
\end{equation}
Using the solution (\ref{4.15}), the integral over $p$ gives a $\delta $%
-function in $r$. With this we can perform the $r$ integration to obtain the
result 
\begin{equation}
P(x;t)=\frac{1}{2\pi \hbar }\int_{-\infty }^{\infty }ds\tilde{W}(Gs,m\dot{G}%
s;0)\exp \{-\frac{1}{2\hbar ^{2}}\left\langle X^{2}\right\rangle s^{2}+i%
\frac{x}{\hbar }s\}.  \label{4.24}
\end{equation}
In the next Section, we evaluate this probability density for some problems
of interest.

\section{Examples}

In this section we consider the evaluation of the general result (\ref{4.24}%
) for the case of a free particle ($K=0$) interacting with an Ohmic bath. We
have chosen the Ohmic model since it is that used most extensively, almost
universally, in discussions of dissipative systems. Newtonian drag is Ohmic,
as is the Stokes force and, of course, classical Brownian motion. In quantum
electrodynamics, the Weisskopf-Wigner approximation is an Ohmic model. In
addition, our results take their simplest form for that model. In Appendix A
we present selected results for the more general single relaxation time
model.

The examples are intended, first of all, to illustrate the power and utility
of our exact solution. They are chosen since they appear in a truly large
number of recent papers, where approximate methods were used. They are also
the examples discussed, by numerical methods, in the original Hu-Paz-Zhang
paper as well as in the earlier papers we have cited.

\subsection{Preliminary formulas}

For the Ohmic bath the memory function has the form 
\begin{equation}
\mu (t)=2\zeta \delta (t),  \label{5.1}
\end{equation}
where $\zeta $ is the Newtonian friction constant (the factor 2 is because
the integral in the Langevin equation is over only half the delta-function).
In this case the equation (\ref{2.10}) is already in local form. With the
form (\ref{5.1}) for the memory function and with $K=0$, the response
function (\ref{2.7}) takes the simple form 
\begin{equation}
\alpha (z)=\frac{1}{-mz^{2}-iz\zeta }.  \label{5.2}
\end{equation}
The Green function (\ref{2.6}) is then 
\begin{equation}
G(t)=\frac{1-e^{-\zeta t/m}}{\zeta }.  \label{5.3}
\end{equation}
If we form the quantities (\ref{2.19}) with this Green function, we find 
\begin{equation}
2\Gamma (t)=\frac{\zeta }{m},\qquad \Omega ^{2}(t)=\frac{2\zeta }{m}\delta
(t).  \label{5.4}
\end{equation}
For any finite time these expressions follow trivially. The delta function
is not so easy to see, although it should be obvious from the form (\ref
{2.10}) of the Langevin equation. In Appendix A, this result is derived
explicitly in the Ohmic limit of the single relaxation time model.

The only other quantity that we need to evaluate the general result is $%
\left\langle X^{2}(t)\right\rangle $, the mean square of the fluctuating
position operator. In this connection, it is a simple matter, comparing the
stationary solution (\ref{2.5}) with the definition (\ref{2.16}) of $X(t)$,
to obtain the general relation 
\begin{equation}
X(t)=x_{{\rm s}}(t)-x_{{\rm s}}(0)+\int_{-\infty }^{0}dt^{\prime
}\{G(-t^{\prime })-G(t-t^{\prime })\}F(t^{\prime }).  \label{5.5}
\end{equation}
In discussing this operator, in particular the mean of its square, we shall
make use of the mean square displacement for the stationary process, 
\begin{eqnarray}
s(t) &\equiv &\left\langle [x_{{\rm s}}(t)-x_{{\rm s}}(0)]^{2}\right\rangle 
\nonumber \\
&=&\frac{2\hbar }{\pi }\int_{0}^{\infty }d\omega {\rm Im}\{\alpha (\omega
+i0^{+})\}\coth \frac{\hbar \omega }{2kT}(1-\cos \omega t).  \label{5.6}
\end{eqnarray}

The discussion in the previous paragraph has been general, applying to an
oscillator interacting with an arbitrary heat bath. We now specialize to the
current case of a free particle interacting with an Ohmic bath. The Green
function is then given by (\ref{5.3}), from which we see that $G(-t^{\prime
})-G(t-t^{\prime })=-mG(t)\dot{G}(-t^{\prime })$ and (\ref{5.5}) becomes 
\begin{equation}
X(t)=x_{{\rm s}}(t)-x_{{\rm s}}(0)-mG(t)\dot{x}_{{\rm s}}(0).  \label{5.7}
\end{equation}
Forming the mean square, we can write 
\begin{equation}
\left\langle X^{2}(t)\right\rangle =s(t)-mG(t)\dot{s}(t)+\frac{1}{2}%
m^{2}G^{2}(t)\ddot{s}(0).  \label{5.8}
\end{equation}
In the same way we find 
\begin{equation}
\left\langle \dot{X}^{2}(t)\right\rangle =\frac{1}{2}[1+m^{2}\dot{G}^{2}]%
\ddot{s}(0)-m\dot{G}(t)\ddot{s}(t).  \label{5.9}
\end{equation}
For this Ohmic case, using the form (\ref{5.2}) for the response function in
the expression (\ref{5.6}) for the mean square displacement, we find 
\begin{equation}
s(t)=\frac{2\hbar \zeta }{\pi }\int_{0}^{\infty }d\omega \frac{\coth \frac{%
\hbar \omega }{2kT}}{\omega (m^{2}\omega ^{2}+\zeta ^{2})}(1-\cos \omega t).
\label{5.10}
\end{equation}

\subsubsection{High temperature limit}

In the high temperature limit ($kT\gg \hbar \zeta /m$) we replace the
hyperbolic cotangent in (\ref{5.10}) by the reciprocal of its argument
(thus, neglecting the zero-point oscillations) The result takes the form 
\cite{bateman} 
\begin{equation}
s(t)=\frac{2kT}{\zeta }(t-m\frac{1-e^{-\zeta t/m}}{\zeta }).  \label{5.11}
\end{equation}
With this in the expression (\ref{5.8}) for the mean square of the
fluctuating displacement, we see that, still in the high temperature limit, 
\begin{equation}
\left\langle X^{2}(t)\right\rangle =\frac{kT}{\zeta }(2t-2m\frac{1-e^{-\zeta
t/m}}{\zeta }-m\frac{(1-e^{-\zeta t/m})^{2}}{\zeta }).  \label{5.12}
\end{equation}

\subsubsection{Zero temperature}

At zero temperature, we replace the hyperbolic cotangent in (\ref{5.10}) by
unity. The result can be written 
\begin{equation}
s(t)=\frac{2\hbar }{\pi \zeta }I(\frac{\zeta t}{m}),  \label{5.13}
\end{equation}
where \cite{bateman} 
\begin{eqnarray}
I(x) &=&\int_{0}^{\infty }dy\frac{x^{2}}{y(y^{2}+x^{2})}(1-\cos y)  \nonumber
\\
&=&\log x+\gamma -\frac{1}{2}[e^{-x}{\rm \bar{E}i}(x)+e^{x}{\rm Ei}(-x)].
\label{5.14}
\end{eqnarray}
Here $\gamma =0.577215665$ is Euler's constant. Note the expansions \cite
{batemanhtf}, for small $x$, 
\begin{equation}
I(x)=-(\log x+\gamma )(\cosh x-1)-\frac{1}{2}[e^{-x}\sum_{n=1}^{\infty }%
\frac{x^{n}}{n!n}+e^{x}\sum_{n=1}^{\infty }\frac{(-x)^{n}}{n!n}],
\label{5.15}
\end{equation}
and, asymptotically, for large $x,$%
\begin{equation}
I(x)\sim \log x+\gamma -\frac{1}{x^{2}}-\frac{3!}{x^{4}}-\frac{5!}{x^{6}}%
-\cdots .  \label{5.16}
\end{equation}
Here we see that there is a serious concern: for small $x$ the second
derivative $I^{\prime \prime }(x)\cong -\log x$ and therefore the term
involving $\ddot{s}(0)$ in the expression (\ref{5.8}) for $\left\langle
X^{2}(t)\right\rangle $ is logarithmically divergent. This divergence
persists for long times, where 
\begin{equation}
\left\langle X^{2}(t)\right\rangle \sim \frac{2\hbar }{\pi \zeta }\log \zeta
t-\frac{\hbar }{\pi \zeta }\log 0^{+},  \label{5.17}
\end{equation}
in which the neglected quantity is of the order of a finite constant. This
divergence for the Ohmic case has, of course, been noted by earlier authors 
\cite{unruh89,hpz}, but it does not seem to be known that this divergence
persists in a model with a high frequency cutoff. In Appendix A we show this
explicitly for the single relaxation time model.

\subsection{Gaussian wave packet}

To begin, we note that in evaluating the solution we make repeated use of
the standard Gaussian integral: 
\begin{equation}
\int_{-\infty }^{\infty }dx\exp \{-\frac{1}{2}ax^{2}+ibx\}=\sqrt{\frac{2\pi 
}{a}}\exp \{-\frac{b^{2}}{2a}\}.  \label{5.18}
\end{equation}
We consider an initial state corresponding to a Gaussian wave packet of the
form, 
\begin{equation}
\psi (x,0)=\frac{1}{(2\pi \sigma ^{2})^{1/4}}\exp \{-\frac{(x-x_{0})^{2}}{%
4\sigma ^{2}}\}.  \label{5.19}
\end{equation}
This is a so-called minimum uncertainty wave packet, centered at $%
\left\langle x(0)\right\rangle =x_{0}$ and with width $\left\langle \Delta
x^{2}(0)\right\rangle \equiv \left\langle \lbrack x(0)-\left\langle
x(0)\right\rangle ]^{2}\right\rangle =\sigma ^{2}$. The corresponding mean
momentum is $\left\langle p(0)\right\rangle =m\left\langle \dot{x}%
(0)\right\rangle =0$ and the momentum width (corresponding to minimum
uncertainty) is $\left\langle \Delta p^{2}(0)\right\rangle
=m^{2}\left\langle [\dot{x}^{2}(0)-\left\langle \dot{x}(0)\right\rangle
]^{2}\right\rangle =\frac{\hbar ^{2}}{4\sigma ^{2}}$. The Wigner function
corresponding to this state is 
\begin{eqnarray}
W(q,p;0) &=&\frac{1}{2\pi \hbar }\int_{-\infty }^{\infty }due^{iup/\hbar
}\psi (q-\frac{u}{2},0)\psi ^{\ast }(q+\frac{u}{2},0)  \nonumber \\
&=&\frac{1}{\pi \hbar }\exp \{-\frac{(q-x_{0})^{2}}{2\sigma ^{2}}-\frac{%
2\sigma ^{2}p^{2}}{\hbar ^{2}}\}.  \label{5.20}
\end{eqnarray}
Its Fourier transform is 
\begin{eqnarray}
\tilde{W}(Q,P;0) &=&\int_{-\infty }^{\infty }dq\int_{-\infty }^{\infty
}dpe^{-i(Pq+Qp)/\hbar }W(q,p;0)  \nonumber \\
&=&\int_{-\infty }^{\infty }dqe^{-iPq/\hbar }\psi (q-\frac{Q}{2},0)\psi
^{\ast }(q+\frac{Q}{2},0)  \nonumber \\
&=&\exp \{-\frac{Q^{2}}{8\sigma ^{2}}-\frac{\sigma ^{2}P^{2}}{2\hbar ^{2}}-i%
\frac{x_{0}P}{\hbar }\}.  \label{5.21}
\end{eqnarray}
Putting this in the expression (\ref{4.24}) for the probability density at
time $t$, we find 
\begin{equation}
P(x;t)=\frac{1}{\sqrt{2\pi \left\langle \Delta x^{2}(t)\right\rangle }}\exp
\{-\frac{[x-m\dot{G}(t)x_{0}]^{2}}{2\left\langle \Delta
x^{2}(t)\right\rangle }\},  \label{5.22}
\end{equation}
where $\left\langle \Delta x^{2}(t)\right\rangle $ is the variance of the
position, given by 
\begin{equation}
\left\langle \Delta x^{2}(t)\right\rangle =m^{2}\dot{G}^{2}(t)\sigma ^{2}+%
\frac{\hbar ^{2}G^{2}(t)}{4\sigma ^{2}}+\left\langle X^{2}(t)\right\rangle .
\label{5.23}
\end{equation}
This is a general result, valid for any heat bath. For the Ohmic case the
Green function is given in (\ref{5.3}) and $\left\langle
X^{2}(t)\right\rangle $ is given by (\ref{5.8}). For the single relaxation
time model, the corresponding results are given in (\ref{A4}) and (\ref{A9})

The first thing that should strike us in this result is that the variance is
in fact infinite, due to the divergence found in the expression (\ref{5.8})
for $\left\langle X^{2}(t)\right\rangle $ at zero temperature. This
divergence arises from the zero-point oscillations and is therefore always
present, although conventionally one neglects the zero-point oscillations in
the high temperature limit. This is a serious difficulty, since it tells us
that an initial wave packet spreads instantly to infinite width. In effect,
the state vanishes! We emphasize that the problem here is not with the
Hu-Paz-Zhang equation or its solution, which are perfectly correct, but with
the assumption of an uncorrelated initial state. Unease with this assumption
has been expressed by many authors (we note in particular the remarks in the
concluding Discussion section of the Hu-Paz-Zhang paper \cite{hpz}) but it
does not seem to have been realized how serious are its consequences.
Indeed, the only meaningful results for the equation are for the high
temperature limit, and we confine our discussion in the following to that
limit.

In the high temperature limit, using the expression (\ref{5.3}) for the
Green function in the Ohmic case, the probability density (\ref{5.22})
becomes 
\begin{equation}
P(x;t)=\frac{1}{\sqrt{2\pi \left\langle \Delta x^{2}(t)\right\rangle }}\exp
\{-\frac{(x-x_{0}e^{-\zeta t/m})^{2}}{2\left\langle \Delta
x^{2}(t)\right\rangle }\},  \label{5.24}
\end{equation}
where, evaluating the expression (\ref{5.8}) for $\left\langle
X^{2}(t)\right\rangle $ with the high-temperature expression (\ref{5.11})
for the mean square displacement, the variance is given by 
\begin{eqnarray}
\left\langle \Delta x^{2}(t)\right\rangle &=&\sigma ^{2}e^{-2\zeta t/m}+%
\frac{\hbar ^{2}(1-e^{-\zeta t/m})^{2}}{4\zeta ^{2}\sigma ^{2}}  \nonumber \\
&&+\frac{kT}{\zeta }(2t-2m\frac{1-e^{-\zeta t/m}}{\zeta }-m\frac{%
(1-e^{-\zeta t/m})^{2}}{\zeta }).  \label{5.25}
\end{eqnarray}

A difficulty with this result is that the center of the wave packet drifts
to the origin. Since for a free particle the origin cannot be a special
point, we see from this that the translational invariance of the problem is
broken by the assumption that the initial state corresponds to an uncoupled
system. Indeed, the system Hamiltonian (\ref{2.9}) is invariant under
simultaneous translation of the particle and bath coordinates ($x\rightarrow
x+d$, $q_{j}\rightarrow q_{j}+d$) and the time-dependent master equation (%
\ref{1.1}) correctly describes the dynamics of the system with regard to
this Hamiltonian. But the bath Hamiltonian (\ref{2.17}) does not possess
this invariance and the initial state is one in which the bath is in
equilibrium with respect to this Hamiltonian. Another way to see this effect
is to note that for a free particle interacting with an Ohmic bath the mean
motion described (\ref{1.1})  satisfies the equation 
\begin{equation}
m\left\langle \ddot{x}\right\rangle +\zeta \left\langle \dot{x}\right\rangle
=-2\zeta \delta (t)\left\langle x(0)\right\rangle .  \label{5.26}
\end{equation}
That is, the particle receives an initial impulse $-\zeta \left\langle
x(0)\right\rangle $, directed toward the origin and with a magnitude such
that in the course of time the particle arrives at the origin. Another
difficulty, which in fact has the same origin as the first, is that in the
expression (\ref{5.25}) for the variance the first term, which corresponds
to the initial variance, decays in time. But the initial variance should
persist and not decay.

A further difficulty is seen if we look at the variance (\ref{5.25}) for
times short compared with the Ohmic decay time , 
\begin{equation}
\left\langle \Delta x^{2}(t)\right\rangle \cong \sigma ^{2}+\frac{\hbar
^{2}t^{2}}{4m^{2}\sigma ^{2}},\qquad \zeta t/m\ll 1.  \label{5.27}
\end{equation}
But this is exactly the formula for the RMS width of a wave packet as
obtained from elementary quantum mechanics \cite{schiff}. It therefore
corresponds to temperature zero and the thermal spreading one should expect
at high temperature is absent\cite{hakim,ford01a,ford01b}. Now this last is
not a difficulty with the equation or our solution, but with the initial
state we have chosen, which corresponds to a particle at temperature zero.
Instead we should choose a state corresponding to a wave packet at
temperature $T$, obtained by averaging the initial Wigner function (\ref
{5.20}) over a thermal distribution of initial velocities.

To accomplish this, we note first that the initial state (\ref{5.19})
corresponds to a particle at rest (i.e., at $T=0$). To obtain the state
corresponding to a particle with velocity $v$, we simply multiply with a
factor $e^{imvx/\hbar }$. With this, we see that the Fourier transform of
the corresponding Wigner function is obtained by multiplying (\ref{5.21}) by 
$e^{-imvQ/\hbar }$. The thermal average of this factor is 
\begin{equation}
\sqrt{\frac{m}{2\pi kT}}\int_{-\infty }^{\infty }dv\exp \{-\frac{mv^{2}}{2kT}%
-i\frac{mQ}{\hbar }v\}=\exp \{-\frac{Q^{2}}{2\bar{\lambda}^{2}}\},
\label{5.28}
\end{equation}
where $\bar{\lambda}$ is the thermal de Broglie wavelength, 
\begin{equation}
\bar{\lambda}=\frac{\hbar }{\sqrt{mkT}}.  \label{5.29}
\end{equation}
Therefore, we see that for a particle at temperature $T$ the Fourier
transform of the initial Wigner function is obtained by multiplying the
corresponding function at $T=0$ by the factor (\ref{5.28}). Multiplying (\ref
{5.21}) by this factor we obtain 
\begin{equation}
\tilde{W}_{T}(Q,P;0)=\exp \{-(\frac{1}{8\sigma ^{2}}+\frac{1}{2\bar{\lambda}%
^{2}})Q^{2}-\frac{\sigma ^{2}P^{2}}{2\hbar ^{2}}-i\frac{x_{0}P}{\hbar }\},
\label{5.30}
\end{equation}
where we indicate that this corresponds to temperature $T$ by the subscript.
Using this in (\ref{4.24}) we find for the probability distribution at
temperature $T$ the result 
\begin{equation}
P_{T}(x;t)=\frac{1}{\sqrt{2\pi \left\langle \Delta x^{2}\right\rangle _{T}}}%
\exp \{-\frac{(x-x_{0}e^{-\zeta t/m})^{2}}{2\left\langle \Delta
x^{2}\right\rangle _{T}}\},  \label{5.31}
\end{equation}
where we have introduced 
\begin{eqnarray}
\left\langle \Delta x^{2}\right\rangle _{T} &=&\left\langle \Delta
x^{2}\right\rangle +\frac{\hbar ^{2}}{\bar{\lambda}^{2}\zeta ^{2}}%
(1-e^{-\zeta t/m})^{2}  \nonumber \\
&=&\sigma ^{2}e^{-2\zeta t/m}+\frac{\hbar ^{2}(1-e^{-\zeta t/m})^{2}}{4\zeta
^{2}\sigma ^{2}}+\left\langle X^{2}(t)\right\rangle _{T},  \label{5.32}
\end{eqnarray}
in which 
\begin{equation}
\left\langle X^{2}(t)\right\rangle _{T}=\frac{kT}{\zeta }(2t-2m\frac{%
1-e^{-\zeta t/m}}{\zeta }).  \label{5.33}
\end{equation}
Note that now the variance for very short times includes the thermal
spreading \cite{ford01a,ford01b}, 
\begin{equation}
\left\langle \Delta x^{2}\right\rangle _{T}\cong \sigma ^{2}+\frac{\hbar
^{2}t^{2}}{4m^{2}\sigma ^{2}}+\frac{kT}{m}t^{2},\qquad \zeta t/m\ll 1.
\label{5.34}
\end{equation}
However, the long time drift of the wave packet center and the shrinking of
the initial variance remain.

\subsection{Pair of Gaussian wave packets}

We consider now an initial state corresponding to two separated Gaussian
wave packets. The corresponding wave function has the form 
\begin{equation}
\psi (x,0)=\frac{1}{(8\pi \sigma ^{2})^{1/4}(1+e^{-\frac{d^{2}}{8\sigma ^{2}}%
})^{1/2}}(\exp \{-\frac{(x-\frac{d}{2})^{2}}{4\sigma ^{2}}\}+\exp \{-\frac{%
(x+\frac{d}{2})^{2}}{4\sigma ^{2}}\}),  \label{5.35}
\end{equation}
where $d$ is the separation and $\sigma $ is the width of each packet. Using
(\ref{5.21}) we see that the Fourier transform of the initial Wigner
function is given by 
\begin{equation}
\tilde{W}(Q,P;0)=\frac{1}{1+e^{-\frac{d^{2}}{8\sigma ^{2}}}}\exp \{-\frac{%
Q^{2}}{8\sigma ^{2}}-\frac{\sigma ^{2}P^{2}}{2\hbar ^{2}}\}(\cos \frac{Pd}{%
2\hbar }+e^{-\frac{d^{2}}{8\sigma ^{2}}}\cosh \frac{Qd}{4\sigma ^{2}}).
\label{5.36}
\end{equation}
Putting this in the expression (\ref{4.24}) for the probability density at
time $t$, we find 
\begin{eqnarray}
P(x;t) &=&\frac{1}{2(1+e^{-d^{2}/2\sigma ^{2}})\sqrt{2\pi \left\langle
\Delta x^{2}\right\rangle }}\left( \exp \{-\frac{(x-\frac{d}{2}e^{-\zeta
t/m})^{2}}{2\left\langle \Delta x^{2}\right\rangle }\}+\exp \{-\frac{(x+%
\frac{d}{2}e^{-\zeta t/m})^{2}}{2\left\langle \Delta x^{2}\right\rangle }%
\}\right.  \nonumber \\
&&\left. +2\exp \{-\frac{x^{2}}{2\left\langle \Delta x^{2}\right\rangle }-%
\frac{d^{2}}{8\sigma ^{2}}+\frac{(1-e^{-\zeta t/m})^{2}\hbar ^{2}d^{2}}{%
32\zeta ^{2}\sigma ^{4}\left\langle \Delta x^{2}\right\rangle }\}\cos \frac{%
G\hbar dx}{4\sigma ^{2}\left\langle \Delta x^{2}\right\rangle }\right) .
\label{5.37}
\end{eqnarray}
Here the first two terms in the parentheses correspond to a pair of Gaussian
wave packets of the form (\ref{5.22}), initially centered at $x=\pm d/2$ and
drifting toward the origin. The third term, that with the cosine, is an
interference term. The attenuation factor $a(t)$ is the ratio of the
coefficient of the cosine term divide by twice the geometric mean of the
first two terms. We find 
\begin{equation}
a(t)=\exp \{-\frac{\left\langle X^{2}(t)\right\rangle }{8\sigma
^{2}\left\langle \Delta x^{2}(t)\right\rangle }d^{2}\},  \label{5.38}
\end{equation}
where at high temperature $\left\langle X^{2}(t)\right\rangle $ is given in (%
\ref{5.12}) and $\left\langle \Delta x^{2}(t)\right\rangle $ in (\ref{5.25}).

For times long compared with the Ohmic decay time, $t\gg m/\zeta $, we see
that $\left\langle X^{2}(t)\right\rangle $ and $\left\langle \Delta
x^{2}\right\rangle $ become asymptotically equal, growing linearly with $t$.
For such long times the attenuation coefficient (\ref{5.38}) therefore
approaches a very small constant, $a(t)\sim \exp \{-\frac{d^{2}}{8\sigma ^{2}%
}\}$. For times short compared with the Ohmic decay time, $t\ll m/\zeta $,
we see that $\left\langle X^{2}(t)\right\rangle \cong \frac{2\zeta kT}{3m^{2}%
}t^{3}$ while $\left\langle \Delta x^{2}(t)\right\rangle \cong \sigma ^{2}+%
\frac{\hbar ^{2}t^{2}}{4m^{2}\sigma ^{2}}$ and therefore 
\begin{equation}
a(t)\cong \exp \{-\frac{\zeta kTd^{2}t^{3}}{12m^{2}\sigma ^{4}+3\hbar
^{2}t^{2}}\},\qquad t\ll m/\zeta .  \label{5.39}
\end{equation}
If we suppose that the slit width is negligibly small, we find $a(t)\cong
\exp \{-t/\tau _{d}\}$ where $\tau _{d}=\frac{3\hbar ^{2}}{\zeta kTd^{2}}$.
This, except for a factor of $6$ is exactly the decoherence time that often
appears in the literature \cite{anglin}. But, as we have seen above, this
result corresponds to a particle in an initial state that is effectively at
temperature zero, which is suddenly coupled to a heat bath at high
temperature. The result is therefore unphysical in the sense that the
initial state does not correspond to that envisioned when we speak of a
system at temperature $T$.

Now, just as in our discussion of the spreading \ of a single Gaussian wave
packet, this difficulty can be repaired by choosing the initial state of the
particle to be the same as that of the heat bath. The prescription for
accomplishing this is very simple: multiply $\tilde{W}(Q,P;0)$, given in (%
\ref{5.36}), by the factor given in (\ref{5.28}). The result is to replace (%
\ref{5.36}) by 
\begin{eqnarray}
\tilde{W}_{{\rm T}}(Q,P;0) &=&\frac{1}{1+e^{-\frac{d^{2}}{8\sigma ^{2}}}}%
\exp \{-(\frac{1}{8\sigma ^{2}}+\frac{1}{2\bar{\lambda}^{2}})Q^{2}-\frac{%
\sigma ^{2}P^{2}}{2\hbar ^{2}}\}  \nonumber \\
&&\times (\cos \frac{Pd}{2\hbar }+e^{-\frac{d^{2}}{8\sigma ^{2}}}\cosh \frac{%
Qd}{4\sigma ^{2}}).  \label{5.40}
\end{eqnarray}
Putting this in the expression (\ref{4.24}) for the probability density at
time $t$, we find, in place of (\ref{5.37}) 
\begin{eqnarray}
P_{{\rm T}}(x;t) &=&\frac{1}{2(1+e^{-d^{2}/2\sigma ^{2}})\sqrt{2\pi
\left\langle \Delta x^{2}\right\rangle _{{\rm T}}}}\left( \exp \{-\frac{(x-%
\frac{d}{2}e^{-\zeta t/m})^{2}}{2\left\langle \Delta x^{2}\right\rangle _{%
{\rm T}}}\}+\exp \{-\frac{(x+\frac{d}{2}e^{-\zeta t/m})^{2}}{2\left\langle
\Delta x^{2}\right\rangle _{{\rm T}}}\}\right.  \nonumber \\
&&\left. +2\exp \{-\frac{x^{2}}{2\left\langle \Delta x^{2}\right\rangle _{%
{\rm T}}}-\frac{d^{2}}{8\sigma ^{2}}+\frac{(1-e^{-\zeta t/m})^{2}\hbar
^{2}d^{2}}{32\zeta ^{2}\sigma ^{4}\left\langle \Delta x^{2}\right\rangle _{%
{\rm T}}}\}\cos \frac{G\hbar dx}{4\sigma ^{2}\left\langle \Delta
x^{2}\right\rangle _{{\rm T}}}\right) ,  \label{5.41}
\end{eqnarray}
where $\left\langle \Delta x^{2}\right\rangle _{{\rm T}}$ is given in (\ref
{5.32}). With this, we find that the attenuation coefficient is given by 
\begin{equation}
a_{{\rm T}}(t)=\exp \{-\frac{\left\langle X^{2}(t)\right\rangle _{{\rm T}}}{%
8\sigma ^{2}\left\langle \Delta x^{2}(t)\right\rangle _{{\rm T}}}d^{2}\}.
\label{5.42}
\end{equation}
Here we recall \ that $\left\langle \Delta x^{2}(t)\right\rangle _{{\rm T}}$
and $\left\langle X^{2}(t)\right\rangle _{{\rm T}}$ are given in (\ref{5.32}%
) and (\ref{5.33}).

Now, for times short compared with the Ohmic decay time we find 
\begin{equation}
a_{{\rm T}}(t)\cong \exp \{-\frac{\frac{kT}{m}t^{2}}{8(\sigma ^{4}+\sigma
^{2}\frac{kT}{m}t^{2}+\frac{\hbar ^{2}}{4m^{2}}t^{2})}d^{2}\},\qquad t\ll
m/\zeta .  \label{5.43}
\end{equation}
This is exactly the form of the attenuation coefficient for a free particle 
\cite{ford01b}, which for very short times is of the form $a_{{\rm T}%
}(t)\cong \exp \{-t^{2}/\tau _{d}^{2}\}$, where the decoherence time is 
\begin{equation}
\tau _{d}=\frac{\sqrt{8}\sigma ^{2}}{\bar{v}d},  \label{5.44}
\end{equation}
in which $\bar{v}=\sqrt{kT/m}$ is the mean thermal velocity.

\section{Concluding remarks}

The system we are discussing is that of an oscillator coupled to a linear
passive heat bath, with a microscopic Hamiltonian of the form (\ref{2.9}).
The long time equilibrium state of this system is entangled, in the sense
that the normal modes correspond to coupled motion of the oscillator and the
bath. The Hu-Paz-Zhang equation is an exact master equation describing how
this entangled equilibrium state arises from an initial state in which the
bath and the oscillator are not coupled. We should perhaps emphasize that
this assumption of a decoupled initial state is common to all derivations of
a master equation, going back at least to the work of Wangness and Bloch 
\cite{wangness}, who phrased it as an assumption that at any instant of time
the system is approximately decoupled. Indeed, such an assumption is
essential for the introduction of the notion of partial trace, i.e., the
trace over states of the uncoupled bath as in (\ref{4.1}), key to the
existence of any master equation. Now, our exact solution has allowed us to
see more clearly how serious is that assumption. In particular, we have seen
that within this assumption an exact solution leads to meaningful results
only in the high temperature limit. Here we hasten to add that this remark
does not apply to the many successful applications of master equations in
weak coupling approximation.

For the most part, previous discussions have been made under the restriction
that the initial state of the particle is a pure state, effectively at zero
temperature, while the bath is at a high temperature $T$. There has even
been an approximate experimental realization of such a state \cite{myatt}.
However, we would argue that such a state is unphysical in the sense that it
does not correspond to what is envisioned when one speaks of a system at
temperature $T$. Rather, the initial time dependence is then dominated by
the ``warming up'' of the particle, which occurs on a\ time scale of order
the decay time $m/\zeta $. On the other hand, as we have shown, the
restriction to such a state is not necessary, one can, within the assumption
of an uncoupled initial state, choose the particle state to be at the same
temperature as the bath.

In order to describe a state of the system that is entangled at all times,
including the initial time, it is necessary to abandon master equation
methods. Some time ago, a more general method applicable to such systems was
described by Ford and Lewis \cite{ford86}. In their method, a system in
equilibrium is put into an initial state (e.g., a wave-packet state) by a
measurement and then at a later time is sampled by a second measurement.
This method of successive measurements has recently been applied to obtain
exact results for the problems of wave packet spreading and decoherence \cite
{ford01a}. For the wave packet spreading one finds in place of (\ref{5.22})
or (\ref{5.31}) the result 
\begin{equation}
P(x;t)=\exp \{-\frac{(x-x_{0})^{2}}{2w^{2}(t)}\},  \label{6.1}
\end{equation}
where the variance is now given by 
\begin{equation}
w^{2}(t)=\sigma ^{2}+s(t)-\frac{[x(t),x(0)]^{2}}{4\sigma ^{2}}.  \label{6.2}
\end{equation}
An equivalent result for wave packet spreading in the Ohmic case has been
obtained by Hakim and Ambegaokar \cite{hakim}, who used functional
integration methods. For the decoherence problem, one obtains in place of (%
\ref{5.38}) or (\ref{5.42}) the result 
\begin{equation}
a(t)=\exp \{-\frac{s(t)}{8\sigma ^{2}w^{2}(t)}\}.  \label{6.3}
\end{equation}
Note that these results are finite at any temperature and apply to an
arbitrary heat bath and for all times. At short times, the results (\ref
{5.34}) and (\ref{5.43}) are in agreement with these exact results.

\appendix

\section{Single relaxation time model}

Here we consider the single relaxation time model for the case of a free
particle ($K=0$). This model corresponds to a memory function of the form 
\begin{equation}
\mu (t)=\frac{\zeta }{\tau }e^{-t/\tau }\theta (t),  \label{A1}
\end{equation}
where $\theta $ is the Heaviside function. Note that in the limit $\tau
\rightarrow 0$ this becomes the Ohmic memory function (\ref{5.1}). With this
form of the memory function and with $K=0$, the response function (\ref{2.7}%
) takes the form 
\begin{eqnarray}
\alpha (z) &=&\frac{1}{-mz^{2}-iz\frac{\zeta }{1-iz\tau }}  \nonumber \\
&=&\frac{z+i(\gamma _{+}+\gamma _{-})}{-mz(z+i\gamma _{+})(z+i\gamma _{-})},
\label{A2}
\end{eqnarray}
where we have introduced 
\begin{equation}
\gamma _{\pm }=\frac{1\pm \sqrt{1-\frac{4\zeta \tau }{m}}}{2\tau }.
\label{A3}
\end{equation}
Note that in the Ohmic limit $\gamma _{+}\rightarrow \tau ^{-1}\rightarrow
\infty \,$\ and $\gamma _{-}\rightarrow \zeta /m$ and we recover the form (%
\ref{5.2}) of the response function.

With this form of the response function, the Green function (\ref{2.6}) can
be written in the form 
\begin{equation}
G(t)=\frac{\gamma _{+}^{2}(1-e^{-\gamma _{-}t})-\gamma _{-}^{2}(1-e^{-\gamma
_{+}t})}{m\gamma _{-}\gamma _{+}(\gamma _{+}-\gamma _{-})},  \label{A4}
\end{equation}
If we form the quantities (\ref{2.19}) with this Green function, we find 
\begin{eqnarray}
2\Gamma (t) &=&\gamma _{-}-\frac{(\gamma _{+}-\gamma _{-})[(\gamma
_{+}+\gamma _{-})e^{\gamma _{-}t}-\gamma _{+}]}{(\gamma _{+}+\gamma
_{-})(e^{\gamma _{+}t}-e^{\gamma _{-}t})+\gamma _{+}-\gamma _{-}},  \nonumber
\\
\Omega ^{2}(t) &=&\frac{\gamma _{-}\gamma _{+}(\gamma _{+}-\gamma _{-})}{%
(\gamma _{+}+\gamma _{-})(e^{\gamma _{+}t}-e^{\gamma _{-}t})+\gamma
_{+}-\gamma _{-}}.  \label{A5}
\end{eqnarray}
In the Ohmic limit, it is clear that for any finite time, $2\Gamma
(t)\rightarrow \gamma _{-}\rightarrow \zeta /m$ and $\Omega
^{2}(t)\rightarrow 0$. On the other hand $2\Gamma (0)=0$ and $\Omega
^{2}(0)=\gamma _{-}\gamma _{+}\rightarrow \frac{\zeta }{m\tau }\rightarrow
\infty $. For $\ t$ of order $\tau $ we see that in this Ohmic limit 
\begin{equation}
\Omega ^{2}(t)\cong \frac{\zeta }{m\tau }e^{-t/\tau }\rightarrow \frac{%
2\zeta }{m}\delta (t).  \label{A6}
\end{equation}
This justifies the assertion made in (\ref{5.4}).

Next, we consider the fluctuating position operator, $X(t)$, for the single
relaxation time model. With the Green function (\ref{A4}) it is a simple
matter to verify the relation 
\begin{equation}
G(-t^{\prime })-G(t-t^{\prime })=-mG(t)\dot{G}(-t^{\prime })-\frac{1-m\dot{G}%
(t)}{\gamma _{+}\gamma _{-}}\ddot{G}(-t^{\prime }).  \label{A7}
\end{equation}
Putting this in (\ref{5.5}) we find in place of (\ref{5.7}), 
\begin{equation}
X(t)=x_{{\rm s}}(t)-x_{{\rm s}}(0)-mG(t)\dot{x}_{{\rm s}}(0)-\frac{m\tau }{%
\zeta }[1-m\dot{G}(t)]\ddot{x}_{{\rm s}}(0).  \label{A8}
\end{equation}
Forming the mean square, we can write 
\begin{eqnarray}
\left\langle X^{2}(t)\right\rangle &=&s(t)-mG(t)\dot{s}(t)+\frac{1}{2}%
m^{2}G^{2}(t)\ddot{s}(0)  \nonumber \\
&&-\frac{m\tau }{\zeta }[1-m\dot{G}(t)][\ddot{s}(t)-\ddot{s}(0)]  \nonumber
\\
&&-\frac{m^{2}\tau ^{2}}{2\zeta ^{2}}[1-m\dot{G}(t)]^{2}s^{(4)}(0).
\label{A9}
\end{eqnarray}
The interest here is in the zero temperature limit. With the response
function given by (\ref{A2}) and with $T=0$, the expression (\ref{5.6}) for
the mean square displacement can be put in the form 
\begin{equation}
s(t)=\frac{2\hbar }{\pi \zeta }\frac{\gamma _{+}^{2}I(\gamma _{-}t)-\gamma
_{-}^{2}I(\gamma _{+}t)}{\gamma _{+}^{2}-\gamma _{-}^{2}},  \label{A10}
\end{equation}
where $I(x)$ is given by (\ref{5.14}). We see now that $\ddot{s}(0)$ is
finite, but the fourth derivative $s^{(4)}(0)$ is logarithmically divergent.
Indeed the divergence is the same as in the Ohmic case, with the same long
time form (\ref{5.17})

\acknowledgments

We wish to thank the School of Theoretical Physics, Dublin Institute for
Advanced Studies, for their hospitality.


\begin{references}
\bibitem[{*}]{bill}  Permanent address: Department of Physics, University of
Michigan, Ann Arbor, Michigan 48109-1120

\bibitem[{{{{{{{{{{{{{{{{{{{{{{{*}}}}}}}}}}}}}}}}}}}}}}}*]{bob}  Permanent
address: Department of Physics and Astronomy, Louisiana State University,
Baton Rouge, Louisiana 70803-4001

\bibitem{hpz}  B. L. Hu, J. P. Paz, and Y. Zhang, Phys. Rev. D {\bf 45},
2843 (1992).

\bibitem{leggett}  A. O. Caldeira and A. J. Leggett, Phys. Rev. A {\bf 31},
1059 (1985)

\bibitem{unruh89}  W. G. Unruh and W. H. Zurek, Phys. Rev. D {\bf 40}, 1071
(1989).

\bibitem{karrlein}  R. Karrlein and H. Grabert, Phys. Rev. E 55, 153 (1997).

\bibitem{halliwell}  J. J. Halliwell and T. Yu, Phys. Rev. D {\bf 53}, 2012
(1996).

\bibitem{ford01a}  G. W. Ford, J. T. Lewis and R. F. O'Connell, Phys. Rev.
A, in press (2001).

\bibitem{ford01b}  G. W. Ford and R. F. O'Connell, Phys. Lett. A, in press
(2001).

\bibitem{flo1988}  G. W. Ford, J. T. Lewis and R. F. O'Connell, Phys. Rev.
37, 4419 (1988).

\bibitem{zwanzig}  R. Zwanzig, J. Stat. Phys. {\bf 3, }215{\bf \ }(1973).

\bibitem{kac}  G. W. Ford and M. Kac, Journ. Stat. Phys. {\bf 45}, 803
(1987).

\bibitem{hillery}  M. Hillery, R. F. O'Connell, M. O. Scully and E. P.
Wigner, Phys. Reports 106, 121 (1984).

\bibitem{landau58}  L. D. Landau and E. M. Lifshitz, {\em Statistical Physics%
} (Addison-Wesley, Reading, MA 1958) page 50.

\bibitem{risken}  H. Risken, {\em The Fokker-Planck equation}, second
edition (Springer Verlag, Berlin 1989).

\bibitem{bateman}  Bateman Manuscript Project, {\em Tables of integral
transforms}, A. Erd\'{e}lyi, Editor, (McGraw-Hill, New York 1954) Vol. 1

\bibitem{batemanhtf}  Bateman Manuscript Project, {\em Higher transcendental
functions}, A. Erd\'{e}lyi, Editor (McGraw-Hill, New York 1954) Vol. 2

\bibitem{schiff}  L. I. Schiff, Quantum mechanics (McGraw-Hill, New York
1949), \ esp. pages 54-59.

\bibitem{hakim}  V. Hakim and V. Ambegaokar, Phys. Rev. A {\bf 32}, 423
(1985).

\bibitem{anglin}  J. R. Anglin, J. P. Paz and W. H. Zurek, Phys. Rev. A {\bf %
55}, 4041 (1997).

\bibitem{wangness}  R. K. Wangness and F. Bloch, Phys. Rev. {\bf 89}, 728
(1953).

\bibitem{myatt}  C. J. Myatt, et al., Nature 403, 269 (2000)

\bibitem{ford86}  G. W. Ford and J. T. Lewis, Probability, Statistical
Mechanics, and Number Theory; Advances in Mathematics Supplemental Studies 
{\bf 9}, 169 (1986).
\end{references}
\end{document}